\documentclass{ws-ijmpd}
\usepackage[compress]{cite}
\usepackage{graphicx}
\usepackage{subfigure}
\usepackage{latexsym}
\usepackage{aas_macros}
\usepackage{mathtools,slashed}
\usepackage[sort&compress,numbers,super]{natbib}
\usepackage{orcidlink}
\usepackage{amssymb, amsmath, bm, physics, color, xcolor}
\usepackage{hyperref}
\hypersetup{
	colorlinks=true,
	linkcolor=blue,
	filecolor=blue,
	urlcolor=blue,
	citecolor=blue,
	pdftitle={Sharelatex Example},
}

\begin{document}

\markboth{Kalita, Govindarajan \& Mukhopadhyay}
{White dwarfs as squashed fuzzy spheres}

\catchline{}{}{}{}{}

\title{Super-Chandrasekhar limiting mass white dwarfs as emergent phenomena of noncommutative squashed fuzzy spheres}

\author{Surajit Kalita\orcidlink{0000-0002-3818-6037}}

\address{Department of Physics, Indian Institute of Science, Bangalore 560012, India\\
	surajitk@iisc.ac.in}

\author{T. R. Govindarajan\orcidlink{0000-0002-8594-0194}}

\address{The Institute of Mathematical Sciences, Chennai 600113, India\\
	trg@imsc.res.in}

\author{Banibrata Mukhopadhyay\orcidlink{0000-0002-3020-9513}}

\address{Department of Physics, Indian Institute of Science, Bangalore 560012, India\\
	bm@iisc.ac.in}

\maketitle

\begin{history}
\received{Day Month Year}
\revised{Day Month Year}
\end{history}

\begin{abstract}
	The indirect evidence for at least a dozen massive white dwarfs violating the Chandrasekhar mass-limit is considered to be one of the wonderful discoveries in astronomy for more than a decade. Researchers have already proposed a diverse amount of models to explain this astounding phenomenon. However, each of these models always carries some drawbacks. On the other hand, noncommutative geometry is one of the best replicas of quantum gravity, which is yet to be proved from observations. Madore introduced the idea of a fuzzy sphere to describe a formalism of noncommutative geometry. This article shows that the idea of a squashed fuzzy sphere can self-consistently explain the super-Chandrasekhar limiting mass white dwarfs. We further show that the length-scale beyond which the noncommutativity is prominent is an emergent phenomenon, and there is no prerequisite for an ad-hoc length-scale.
	
	\keywords{Noncommutative geometry; General relativity; White dwarfs; Landau levels.}
\end{abstract}

\ccode{PACS Nos.: 02.40.Gh, 97.20.Rp, 04.20.-q, 97.10.Nf, 71.70.Di}

\section{Introduction}\label{Introduction}

The end phase of a star possessing a mass $M\lesssim (10\pm 2) \rm\,M_\odot$ \cite{2018MNRAS.480.1547L} is a white dwarf (WD), where the inward pressure due to gravity balances the outward pressure due to the degenerate electron gas, to maintain its stable equilibrium \cite{compact}. Chandrasekhar first proposed thereby a limiting mass of the WD and estimated the maximum mass of a stable non-rotating non-magnetized carbon-oxygen WD to be $\sim 1.4 \rm\,M_\odot$ \cite{1931ApJ....74...81C}. Beyond this mass-limit, the pressure balance no longer sustains, and the WDs explode due to new fusion reactions to produce type Ia supernovae (SNe\,Ia). However, recent observations from more than a dozen of over-luminous SNe\,Ia suggest that the maximum mass of a WD could be as high as $\sim 2.8 \rm\,M_\odot$ \cite{2006Natur.443..308H,2007ApJ...669L..17H,2009ApJ...707L.118Y,2010ApJ...715.1338Y,2010ApJ...714.1209T,2010ApJ...713.1073S,2011MNRAS.410..585S,2011MNRAS.412.2735T,2012ApJ...757...12S}, indicating a clear violation of the Chandrasekhar mass-limit. Moreover, various simulations have already ruled out the possibility for the existence of a double degenerate scenario for the generation of the over-luminous SNe\,Ia, and $2.8\rm\,M_\odot$ progenitor mass as such a double degenerate WD produces an off-center ignition and forms a neutron star rather than an over-luminous SN\,Ia \cite{2004ApJ...615..444S,2006MNRAS.373..263M}. All these indirect evidences suggest that conventional physics may significantly be violated inside the high-density regime of massive WDs.

Ostriker and Hartwick first showed that the rotation could alone increase the mass of a WD up to $\sim1.8\rm\,M_\odot$ \cite{1968ApJ...153..797O}, which, however, cannot explain the inferred significantly super-Chandrasekhar WDs. Thereafter, Mukhopadhyay and his collaborators showed that the high constant magnetic field or highly fluctuating magnetic field such that the average field is low could comprise super-Chandrasekhar WDs with mass up-to $\sim2.6\rm\,M_\odot$ due to the formation of Landau levels \cite{2012MPLA...2750084K,2013PhRvL.110g1102D}. However, such a high field may also destabilize the WD, if it is varying with radial coordinate, depending on the field geometry, as the magnetic field can change its shape and size (which is a classical effect of the magnetic field) \cite{2015MNRAS.454..752S,2019MNRAS.490.2692K,2020IAUS..357...79K,2020ApJ...896...69K}. As a result, having a high magnetic field inside the WDs is a question of debate over the years. Later, various researchers have proposed different models, such as modified gravity \cite{2018JCAP...09..007K,2017EPJC...77..871C}, presence of charge in a WD \cite{2014PhRvD..89j4043L}, generalized Heisenberg uncertainty principle \cite{2018JCAP...09..015O}, lepton number violation \cite{2015NuPhA.937...17B}, and many more to explain the super-Chandrasekhar WDs. We have earlier shown that in the presence of noncommutativity (NC), high-density WDs can have a mass up to $\sim 2.6 \rm\,M_\odot$ \cite{2021IJMPD..3050034K}. While deriving this mass-limit, we proposed based on Wigner's idea \cite{1958PhRv..109..571S} that the effect of NC is significant, mostly at the center of a dense WD, where the inter-electron separation is less than the Compton wavelength.

Noncommutative geometry is a sub-class of quantum gravity that can explain various interesting phenomena ranging from the early universe cosmology to the inside of the event horizon of a black hole. However, there is no direct observational evidence for NC so far. In ordinary quantum mechanics (QM), the position variable $(\hat{x})$ does not commute with the corresponding momentum component $(\hat{p})$ and follows the Heisenberg's uncertainty principle, whereas various components of $\hat{x}$ and $\hat{p}$ commute among themselves. In noncommutative geometry, $\hat{x}$ and $\hat{p}$ even do not commute among themselves. One of the popular ways of proposing NC is just an ad-hoc consideration of $\comm{\hat{x}_i}{\hat{x}_j} = i\eta$ and $\comm{\hat{p}_i}{\hat{p}_j} = i\theta$, where $\eta$ and $\theta$ are the noncommutative variables. Nicolini et al. \cite{2006PhLB..632..547N} used this form of NC to show the shift of the event horizon for black holes. They also showed that this form of NC removes the essential singularity present at the black hole center. Similarly, other researchers have used different forms of noncommutative geometry to describe various problems of physics related to the fundamental length-scale, Berry curvature, Landau levels, etc. \cite{1999JHEP...09..032S,2001PhLB..510..255A,2002PhRvL..88s0403M,2002Natur.418...34A,2006LNP...698...97N,2012PhRvL.109r1602S}. As mentioned above, for many of these theories, the basic assumption in the structure of NC among the position and momentum variables is quite ad-hoc. Madore \cite{1992CQGra...9...69M} first introduced a self-consistent model of $3$-dimensional NC, named fuzzy sphere, where the position variables follow the well-known angular momentum algebra of QM. This formalism was used by various researchers to obtain the thermodynamical properties of non-interacting degenerate electron gas \cite{2014JPhA...47R5203C,2015PhRvD..92l5013S}. Andronache and Steinacker \cite{2015JPhA...48C5401A} later modified the idea of a fuzzy sphere by projecting it on an equatorial plane, named this configuration a squashed fuzzy sphere, and showed that this model of NC mimics the magnetic field in the non-relativistic limit. They also reckoned that this form of NC could quantize the plane by forming different energy levels in the semi-classical limit, which is similar to the Landau levels formed in the presence of a magnetic field in the non-relativistic limit.

In this paper, we use the idea of a squashed fuzzy sphere to apply it in a WD. This formalism is superior to the earlier one we used in \cite{2021IJMPD..3050034K} due to two main reasons. First, fuzzy sphere algebra is symmetric under rotation, and second, the squashed fuzzy sphere captures naturally expected commutative geometry at the boundary of the WD as the NC decreases from the center to the surface. We show that, by assuming the WD as a noncommutative fuzzy sphere, its mass increases significantly to form a super-Chandrasekhar limiting mass WD. Moreover, in this formalism, the length-scale of the system over which effect of NC is prominent depends on the Compton wavelength of an electron and the Planck scale; thereby, it turns out to be an emergent phenomenon. There is no need to pre-assume a length-scale for the prominence of NC. In the presence of NC, we show that fermions behave less like fermions, and hence, the effective pressure between them reduces, which allows more mass to accumulate.

The plan of this paper is as follows. In \S \ref{Squashed fuzzy sphere}, we discuss the basic formalism of the squashed fuzzy sphere and thereby obtain the exact (i.e., applicable in both relativistic and non-relativistic regimes) energy dispersion relation for the squashed fuzzy sphere in \S \ref{Energy dispersion relation}. We compare this dispersion relation with the well-known energy dispersion relation for the Landau levels formed due to a magnetic field. Thereafter, in \S \ref{Degenerate equation of state in squashed fuzzy sphere}, we derive the equation of state (EoS) for the degenerate electrons in a squashed fuzzy sphere. In this section, we also discuss the emergence of an effective length-scale for the prominence of NC. Further, we use the degenerate equation of state in \S \ref{Limiting mass of the WD in squashed fuzzy sphere} to derive the mass--radius relation for the WD, and thereby to obtain its limiting mass. Moreover, we use the dispersion relation to discuss the neutron drip briefly in the presence of NC in \S \ref{Neutron drip in presence of noncommutativity} and to show the region on which this EoS is valid. Finally we end with conclusions in \S \ref{Conclusions}.

\section{Squashed fuzzy sphere}\label{Squashed fuzzy sphere}

A fuzzy sphere $S_N^2$, first introduced by Madore \cite{1992CQGra...9...69M}, is like an ordinary sphere $S^2$, except that its coordinates follow the angular momentum algebra of usual QM and, hence, they, in general, do not commute among themselves. In $\mathbb{R}^3$, a sphere of radius $r \in \mathbb{R}$ is defined as the set of points which follows
\begin{equation}\label{Eq: sphere}
X_1^2 + X_2^2 + X_3^2 = r^2,
\end{equation}
with $X_1$, $X_2$, $X_3$ being the ordinary cartesian coordinates of the points. In a fuzzy sphere, the coordinates $X_i$ ($i=1,2,3$) follows 
\begin{equation}
X_i = \kappa J_i,
\end{equation}
where $\kappa$ is the proportionality (scaling) constant and $J_i$ are the generators of $\mathtt{SU}(2)$ group which follows the angular momentum algebra in an $N$-dimensional irreducible representation such that
\begin{equation}
J_1^2 + J_2^2 + J_3^2 = \frac{\hbar^2}{4}\left(N^2-1\right) \mathbb{I} = C_N \mathbb{I},
\end{equation}
with $C_N = \hbar^2\left(N^2-1\right)/4$, $\hbar = h/2\pi$, $h$ is the Planck constant and $\mathbb{I}$ is the $N$-dimensional identity matrix. Substituting $J_i$ in terms of $X_i$ and defining $k = \kappa r$, we obtain the following relations
\begin{align}\label{Eq: k_R relation}
\kappa = \frac{r}{\sqrt{C_N}} \quad \mathrm{and} \quad k = \frac{r^2}{\sqrt{C_N}}.
\end{align}
Since the angular momentum algebra follows $\comm{J_i}{J_j} = i\hbar\epsilon_{ijk}J_k$, the coordinates of the fuzzy sphere needs to obey the following commutation relation
\begin{equation}\label{Eq: Fuzzy sphere}
\comm{X_i}{X_j} = i\frac{k \hbar}{r} \epsilon_{ijk} X_k.
\end{equation}
Moreover, in a fuzzy sphere, the fuzzy Laplacian is defined as \cite{2015JPhA...48C5401A}
\begin{equation}
\Box = \frac{1}{k^2}\sum_{i=1}^3 \comm{X_i}{\comm{X_i}{.}},
\end{equation}
which satisfies the following eigenvalue equation
\begin{equation}
\Box \hat{Y}^l_m = \frac{\hbar^2}{r^2} l(l+1) \hat{Y}^l_m,
\end{equation}
where $l(l+1)$ are the eigenvalues and $\hat{Y}^l_m$ are the eigenfunctions of the fuzzy Laplacian with $l$ taking all the integer values from $0$ to $N-1$ and $m$ taking all the integer values from $-l$ to $l$.

\begin{figure}[htpb]
	\centering
	\includegraphics[scale=0.4]{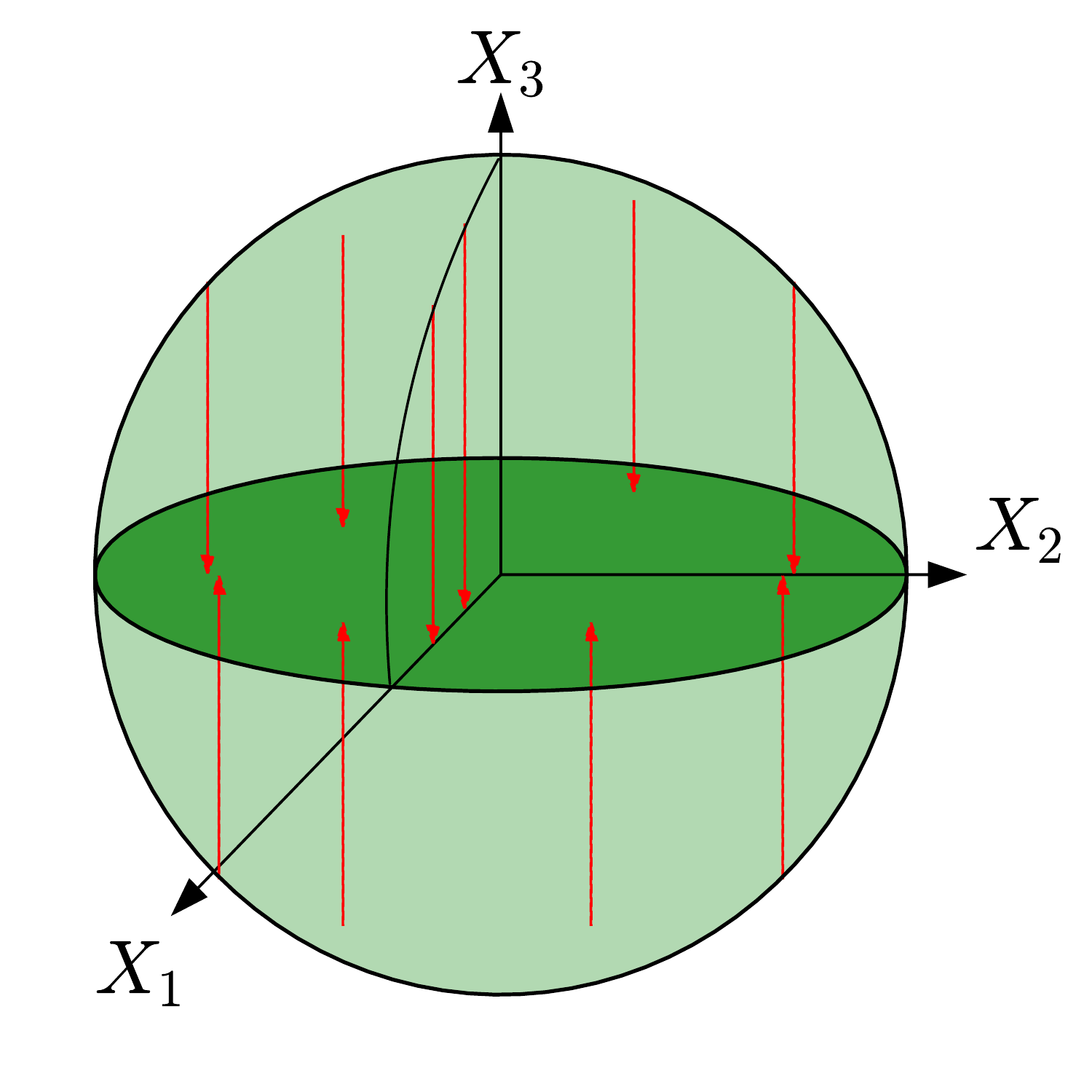}
	\caption{A diagram of a squashed fuzzy sphere which is obtained by projecting all the points of the fuzzy sphere on an equatorial plane. For illustration, $X_1$-$X_2$ is the squashing plane.}
	\label{Fig: squashed fuzzy sphere}
\end{figure}
A squashed fuzzy sphere is interpreted as the configuration when all the points of a fuzzy sphere are projected on any of its equatorial planes. Note that this projection is not a stereographic projection. Figure \ref{Fig: squashed fuzzy sphere} shows the projection of all the points of a fuzzy sphere on the $X_1$-$X_2$ equatorial plane. The upper hemisphere points are projected on the upper side of the equatorial plane and, the points of the lower hemisphere are projected on its lower side and, then, they are glued together. Writing $X_3$ in terms of $X_1$ and $X_2$ from Equation~\eqref{Eq: sphere}, and substituting it in Equation~\eqref{Eq: Fuzzy sphere}, the commutation relation for a squashed fuzzy sphere, with squashing in $X_1$-$X_2$ plane, is modified as \cite{2015JPhA...48C5401A}
\begin{equation}\label{Eq: Squashed fuzzy sphere}
\comm{X_1}{X_2} = \pm i \frac{k \hbar}{r} \sqrt{r^2 - X_1^2 - X_2^2}.
\end{equation}
The Laplacian for the squashed fuzzy sphere is given by \cite{2015JPhA...48C5401A}
\begin{equation}
\Box_s = \frac{1}{k^2}\sum_{i=1}^2 \comm{X_i}{\comm{X_i}{.}},
\end{equation}
which satisfies the following eigenvalue equation
\begin{equation}
\Box_s \hat{Y}^l_m = \frac{\hbar^2}{r^2}\left\{l(l+1)-m^2 \right\}\hat{Y}^l_m,
\end{equation}
where $l(l+1)-m^2$ are eigenvalues of the squashed fuzzy Laplacian.

\section{Energy dispersion relation in a squashed fuzzy sphere}\label{Energy dispersion relation}

\begin{figure}[!htbp]
	\centering
	\subfigure[]{\includegraphics[scale = 0.38]{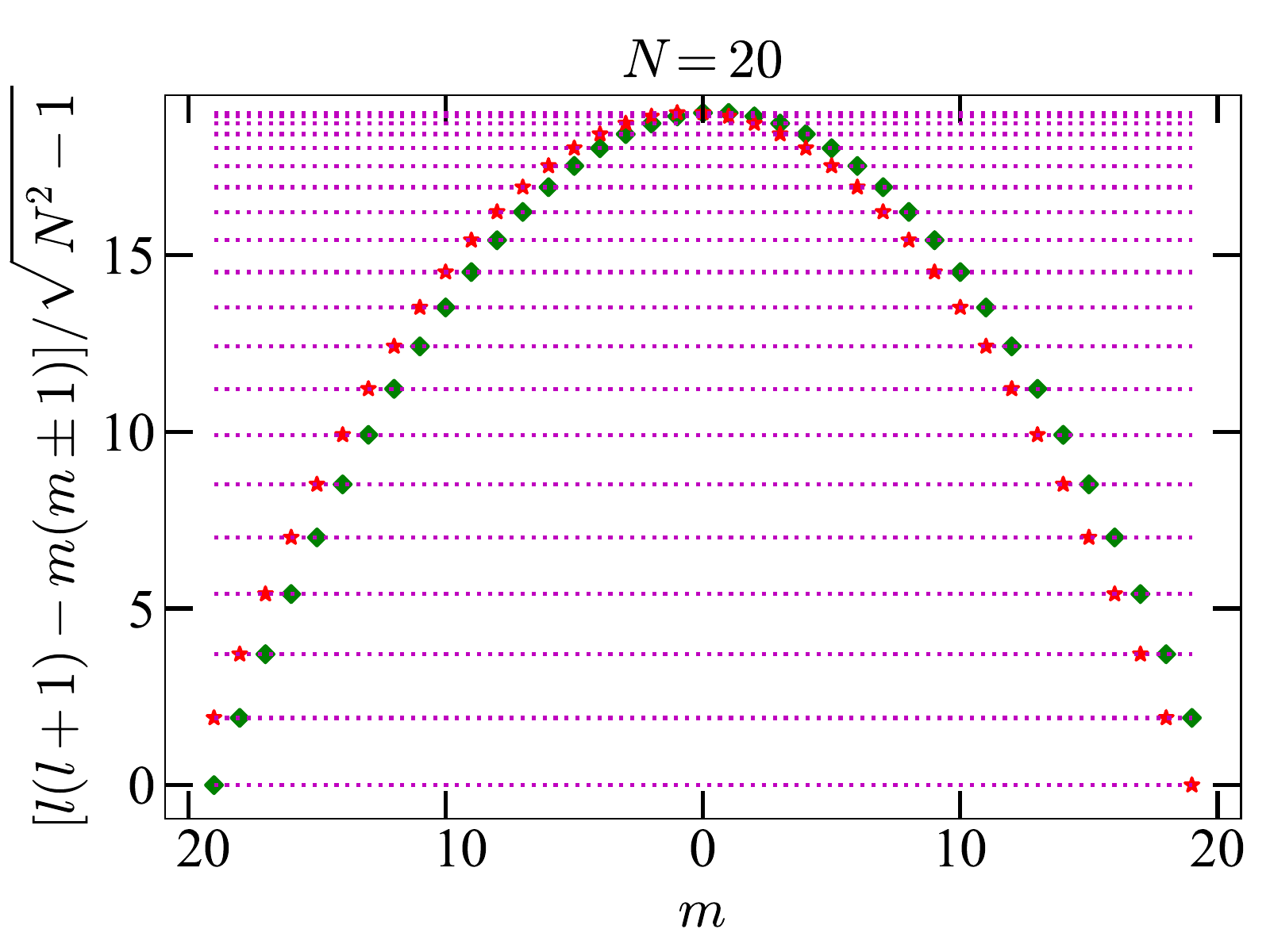}}
	\subfigure[]{\includegraphics[scale = 0.38]{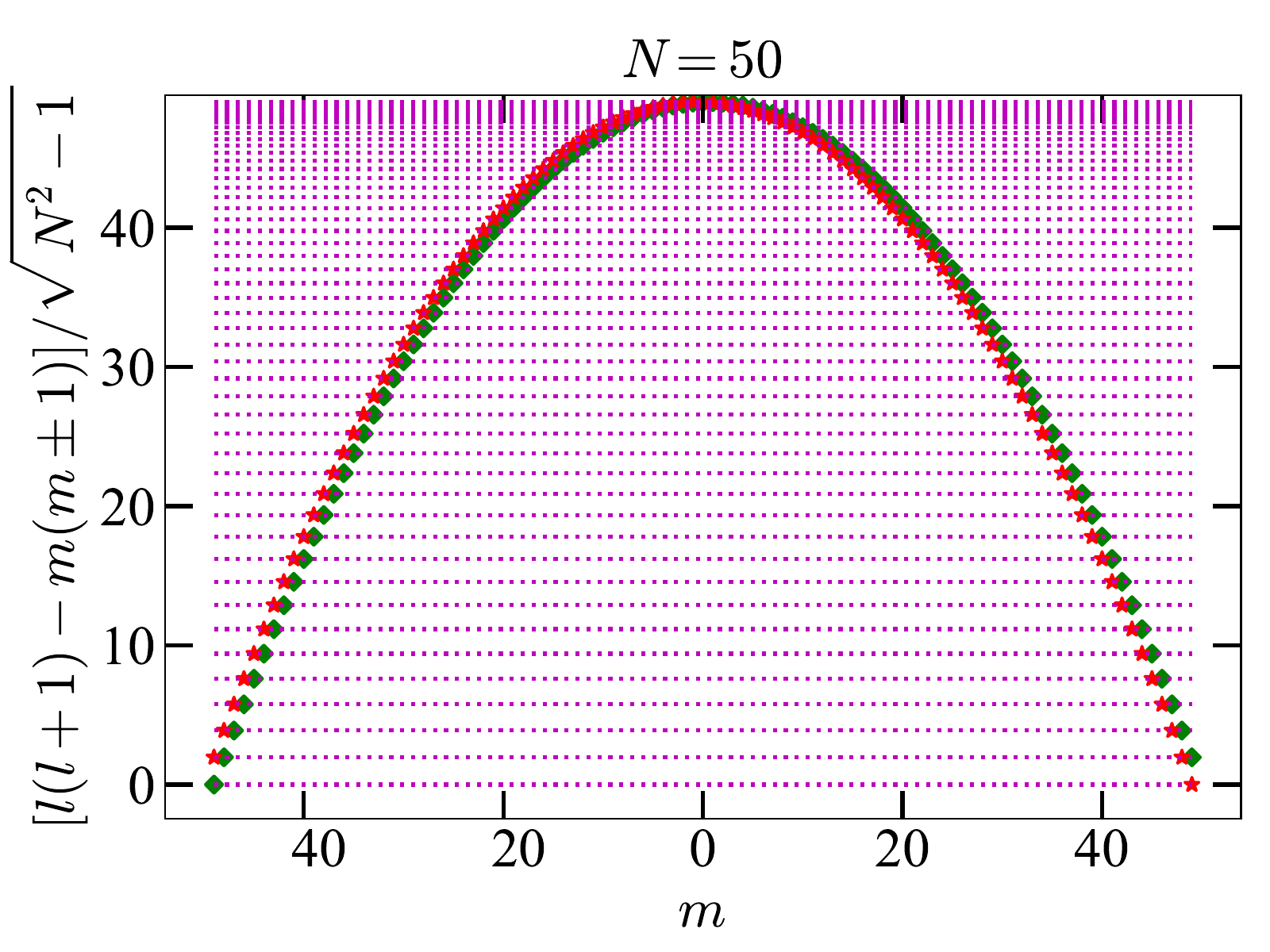}}
	\caption{Quantized energy levels in the noncommutative squashed fuzzy sphere. Here red stars represent the spin-up electrons and the green squares depict the spin-down electrons. It is evident that the ground level is occupied by one pair of spin-up and spin-down electrons and all the other higher energy levels have two pairs of them. Moreover, the energy gap decreases towards the higher energy levels.}
	\label{Fig: Total energy levels}
\end{figure}

\begin{figure}[!htbp]
	\centering
	\subfigure[]{\includegraphics[scale = 0.38]{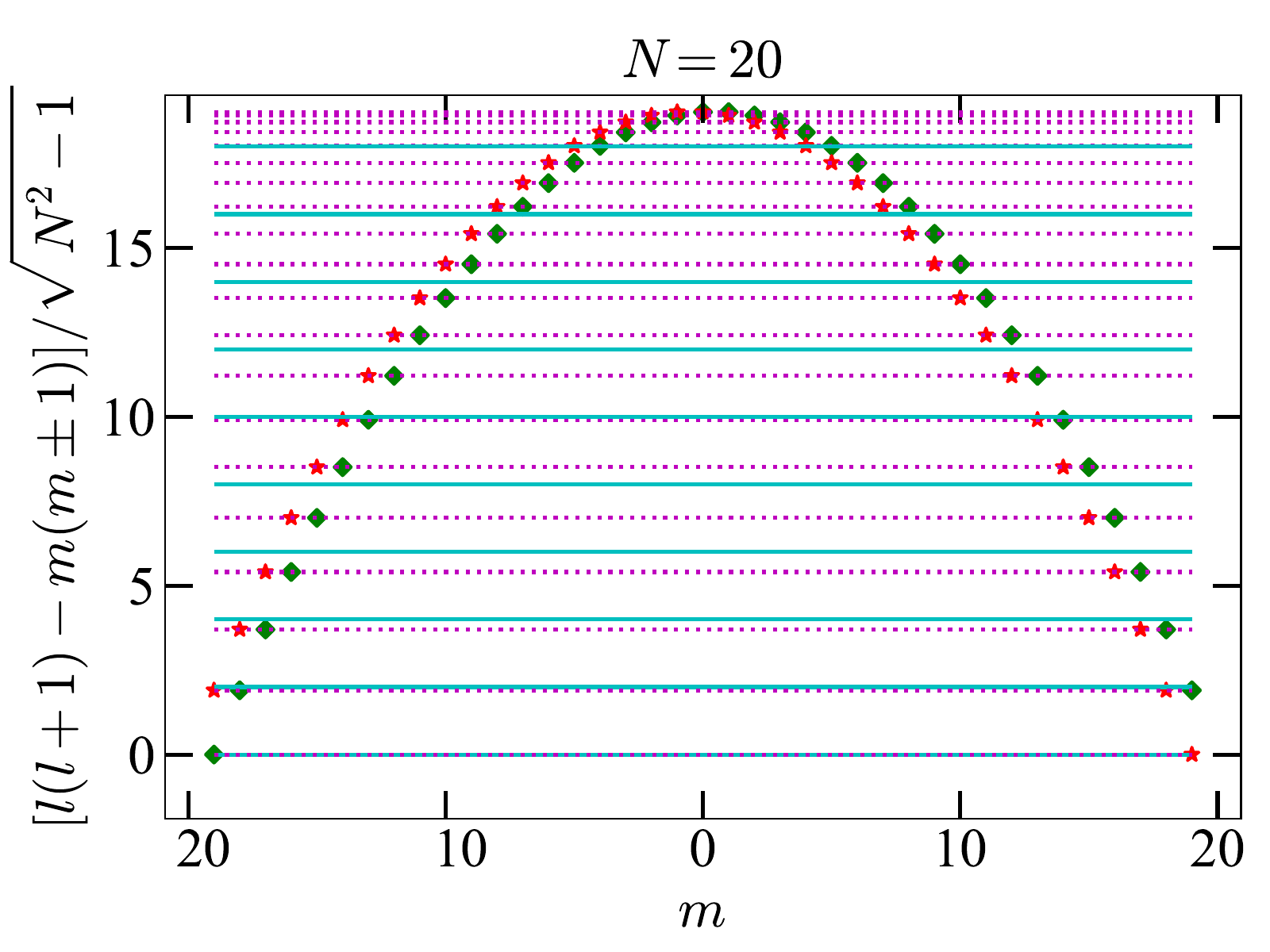}}
	\subfigure[]{\includegraphics[scale = 0.38]{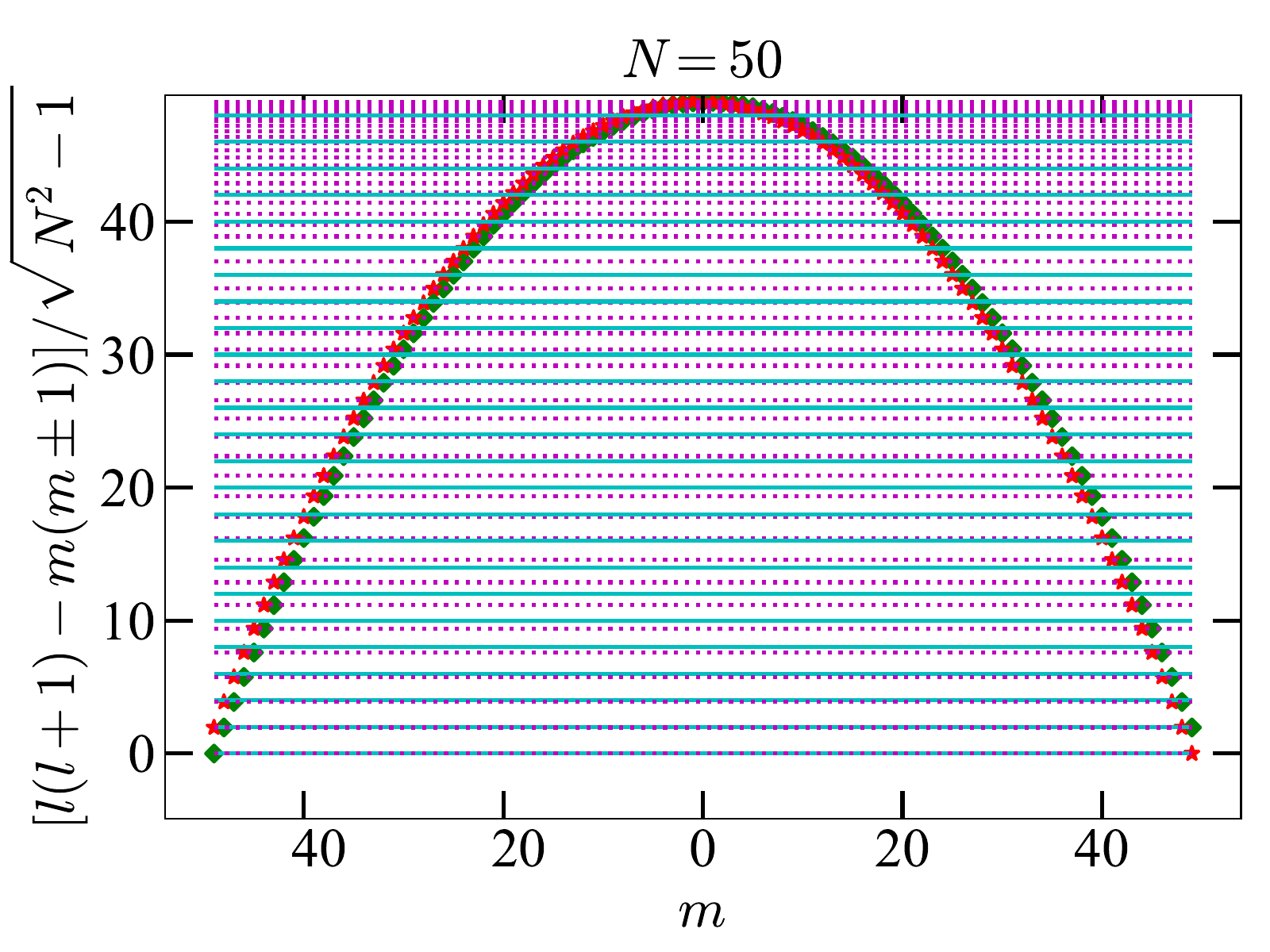}}
	\caption{Comparison of the energy levels obtained from equations~\eqref{Eq: Fuzzy dispersion relation: v1} and \eqref{Eq: Fuzzy dispersion relation}. Here the dotted magenta lines represent the exact energy levels, and the solid cyan lines show the approximated energy levels. It is evident that as $N$ increases, more energy levels coincide with approximated energy levels. Here for $N=20$, two levels coincide, whereas, for $N=50$, five energy levels coincide.}
	\label{Fig: Compare energy levels}
\end{figure}

The Dirac operator for the squashed fuzzy sphere is defined as \cite{2015JPhA...48C5401A}
\begin{equation}\label{Eq: Dirac operator}
\slashed{D} = \frac{c}{k} \left(\sigma_1 \otimes \comm{X_1}{.} + \sigma_2 \otimes \comm{X_2}{.} \right),
\end{equation}
where $c$ is the speed of light and $\sigma_1, \sigma_2, \sigma_3$ are the Pauli spin matrices. The eigenvalues of this Dirac operator gives the quantized energy in a squashed fuzzy sphere. Now, the square of the Dirac operator is given by
\begin{align}
\slashed{D}^2 &= \frac{c^2}{k^2} \sigma_i \sigma_j \otimes \comm{X_i}{\comm{X_i}{.}}\\
&= c^2\left \{ \Box\otimes \mathbb{I}_2 - \frac{1}{k^2} \left( \comm{X_3}{.} + \frac{k \hbar}{2r} \sigma_3\right)^2 + \frac{\hbar^2}{4r^2} \right \}.
\end{align}
Hence, the eigenvalues of $\slashed{D}^2$ (which are the squares of energies) are given by
\begin{align}
E_{l,m}^2 = \frac{\hbar^2 c^2}{r^2} \left\{l(l+1)-\left(m \pm \frac{1}{2}\right)^2 + \frac{1}{4}\right\}.
\end{align}
Since $\sigma_3$ has the eigenvalues $\pm1$, hence the $+$ and $-$ correspond to the energies of the spin-up and spin-down particles respectively. From Equation~\eqref{Eq: k_R relation}, using the relation $k \hbar = 2r^2/\sqrt{N^2-1}$, the above equation reduces to
\begin{align} \label{Eq: enrgy eigenvalues}
E_{l,m}^2 = \frac{2 \hbar c^2}{k \sqrt{N^2-1}} \left\{l(l+1)-m(m\pm 1)\right\}.
\end{align}
Moreover, the relations between the Cartesian and spherical polar coordinates are $X_1=r \sin\theta \cos\phi$ and $X_2=r \sin\theta \sin\phi$ with radius $r$ being fixed for a particular fuzzy sphere. Therefore, using Equation~\eqref{Eq: Squashed fuzzy sphere}, the algebra of a squashed fuzzy sphere in spherical polar coordinates can be recast as
\begin{align}
\comm{\sin\theta \cos\phi}{\sin\theta \sin\phi} &= \pm i \frac{k \hbar}{r^2} \cos\theta.
\end{align}
It is evident that the NC in spherical coordinates is between $\theta$- and $\phi$-coordinates only, while the $r$-coordinate remains free. In other words, the formalism of squashed fuzzy sphere is such that it actually provides a NC between its azimuthal and polar coordinates because the squashed plane in a fuzzy sphere can be any of its equatorial planes and there is no particular direction for it, which means that the squashed fuzzy sphere has a rotational symmetry about the equatorial plane. It can also be verified from Equation~\eqref{Eq: Dirac operator}, where it is evident that we do not require only $\sigma_1$ and $\sigma_2$ to define the Dirac operator and it can be any two of $\sigma_1$, $\sigma_2$, and $\sigma_3$ or their suitable linear combinations. Since all the Pauli matrices have eigenvalues $\pm 1$, the energy expression in Equation~\eqref{Eq: enrgy eigenvalues} will not alter, which also indicates that the squashing has a rotational symmetry and it does not matter which plane we consider. It further indicates that the NC is indeed between $\theta$- and $\phi$-coordinates. Hence, an electron with mass $m_\mathrm{e}$, moving in a squashed fuzzy sphere, does not experience the effect of NC along $r$-coordinate. Therefore, defining $N = l_\mathrm{max}+1$, $\tilde{L} = l_\mathrm{max}-l$, $\tilde{M} = l-m$ and $\tilde{M}'=l+m$, the exact energy dispersion relation for an electron, moving in a squashed fuzzy sphere, is given by
\begin{align}\label{Eq: Fuzzy dispersion relation: v1}
E^2 &= p_r^2 c^2 + m_\mathrm{e}^2 c^4 \left[1+ \{l(l+1)-m(m \pm 1)\} \frac{\hbar^2}{m_\mathrm{e}^2 c^2 r^2}\right] \\
&= p_r^2 c^2 + m_\mathrm{e}^2 c^4 \left[1+ \{l(l+1)-m(m \pm 1)\} \frac{2 \hbar}{m_\mathrm{e}^2 c^2 k \sqrt{N^2-1}}\right] \\
&= \begin{cases}
p_r^2 c^2 + m_\mathrm{e}^2 c^4 \left[1+ \tilde{M} \frac{2N-2\tilde{L}-\tilde{M}-1}{\sqrt{N^2-1}} \frac{2 \hbar}{m_\mathrm{e}^2 c^2 k}\right] \quad \text{for spin-up electrons}\\
p_r^2 c^2 + m_\mathrm{e}^2 c^4 \left[1+ \tilde{M}' \frac{2N-2\tilde{L}-\tilde{M}'-1}{\sqrt{N^2-1}} \frac{2 \hbar}{m_\mathrm{e}^2 c^2 k}\right] \quad \text{for spin-down electrons},
\end{cases} \label{Eq: Fuzzy dispersion relation: v2}
\end{align}
where $p_r$ is the momentum of the electron in radial direction. From Equation~\eqref{Eq: Fuzzy dispersion relation: v1}, it is evident that for spin-up electrons, $m=l$ represents the ground level, whereas $m=l-1$ as well as $m=-l$ are the first energy levels, and so on. Similarly, for spin-down electrons, $m=-l$ is the ground level, whereas $m=-l+1$ and $m=l$ are the first energy levels. Hence, the ground level has always one pair of spin-up and spin-down electrons, whereas all the other energy levels comprise a couple of pairs of spin-up and spin-down electrons. 
Note that `spin-up' and `spin-down' are just the nomenclature to represent respectively the $+$ and $-$ eigenvalues of the Pauli spin matrices. Note that for the present system under consideration, there is no external magnetic field.
Figure \ref{Fig: Total energy levels} shows the quantized energy levels in a noncommutative squashed fuzzy sphere for $N=20$ and $N=50$. From this figure, it is evident that the spin-up and spin-down electrons have same energies and the spacing between the energy levels decreases towards the higher energy levels.
Moreover, in the large $N$ limit, the expression for energy dispersion relation \eqref{Eq: Fuzzy dispersion relation: v2} reduces to
\begin{align}\label{Eq: Fuzzy dispersion relation}
E^2 = p_r^2 c^2 + m_\mathrm{e}^2 c^4 \left(1+ 2\nu \frac{2 \hbar}{m_\mathrm{e}^2 c^2 k}\right),\qquad \nu \in \mathbb{Z}^{0+}
\end{align}
with $\nu = \tilde{M}$ for for spin-up and $\nu=\tilde{M}'$ for for spin-down electrons. This is the quantized energy for an electron in a squashed fuzzy sphere. Comparing it with the energy dispersion relation of Landau quantization due to the magnetic field, which is \cite{1991ApJ...383..745L}
\begin{align}\label{Eq: Magnetic EoS}
E^2 &= p_z^2c^2 + m_\mathrm{e}^2 c^4 \left(1+ 2\nu \frac{B}{B_\mathrm{c}}\right),\qquad \nu \in \mathbb{Z}^{0+}
\end{align}
where $p_z$ is the momentum of an electron along $z$-direction, $B$ is the magnetic field along $z$-direction, and $B_\mathrm{c}$ is the critical magnetic field (Schwinger limit), given by $B_\mathrm{c} = m_\mathrm{e}^2 c^3/\hbar e$ with $e$ being the charge of an electron. Upon simplification, comparing equations~\eqref{Eq: Fuzzy dispersion relation} and \eqref{Eq: Magnetic EoS}, we obtain
\begin{align}
B \equiv \frac{2c}{ek}.
\end{align}
From the above expression, it is evident that the term $k^{-1}$ behaves as NC strength in a squashed fuzzy sphere. Andronache and Steinacker earlier obtained a similar relation between $B$ and $k$ in the case of non-relativistic squashed fuzzy sphere in NC \cite{2015JPhA...48C5401A}. Figure \ref{Fig: Compare energy levels} shows the comparison of the approximated form of dispersion relation of Equation~\eqref{Eq: Fuzzy dispersion relation} with the exact expression from Equation~\eqref{Eq: Fuzzy dispersion relation: v1} for $N=20$ and $N=50$. Comparing both the figures, it is evident that the more number of levels coincide when $N$ is large. For $N=20$, only the ground level and the first level match with the approximated levels, whereas, for $N=50$, five levels coincide with the approximated levels. This shows a good agreement between equations~\eqref{Eq: Fuzzy dispersion relation: v1} and \eqref{Eq: Fuzzy dispersion relation} if $N\ggg1$. Nevertheless, there is a difference with the case of Landau quantization in the presence of magnetic field. In the case of magnetic field, the ground level is singly degenerate and all the other levels are doubly degenerate, whereas in the case of NC, the ground level is doubly degenerate and all the higher levels are quadruply degenerate. Hence, in NC, `spin-up' and `spin-down' electrons have same energies, which is not the case in ground level for magnetic Landau quantization.

\begin{figure}[htpb]
	\centering
	\includegraphics[scale=0.4]{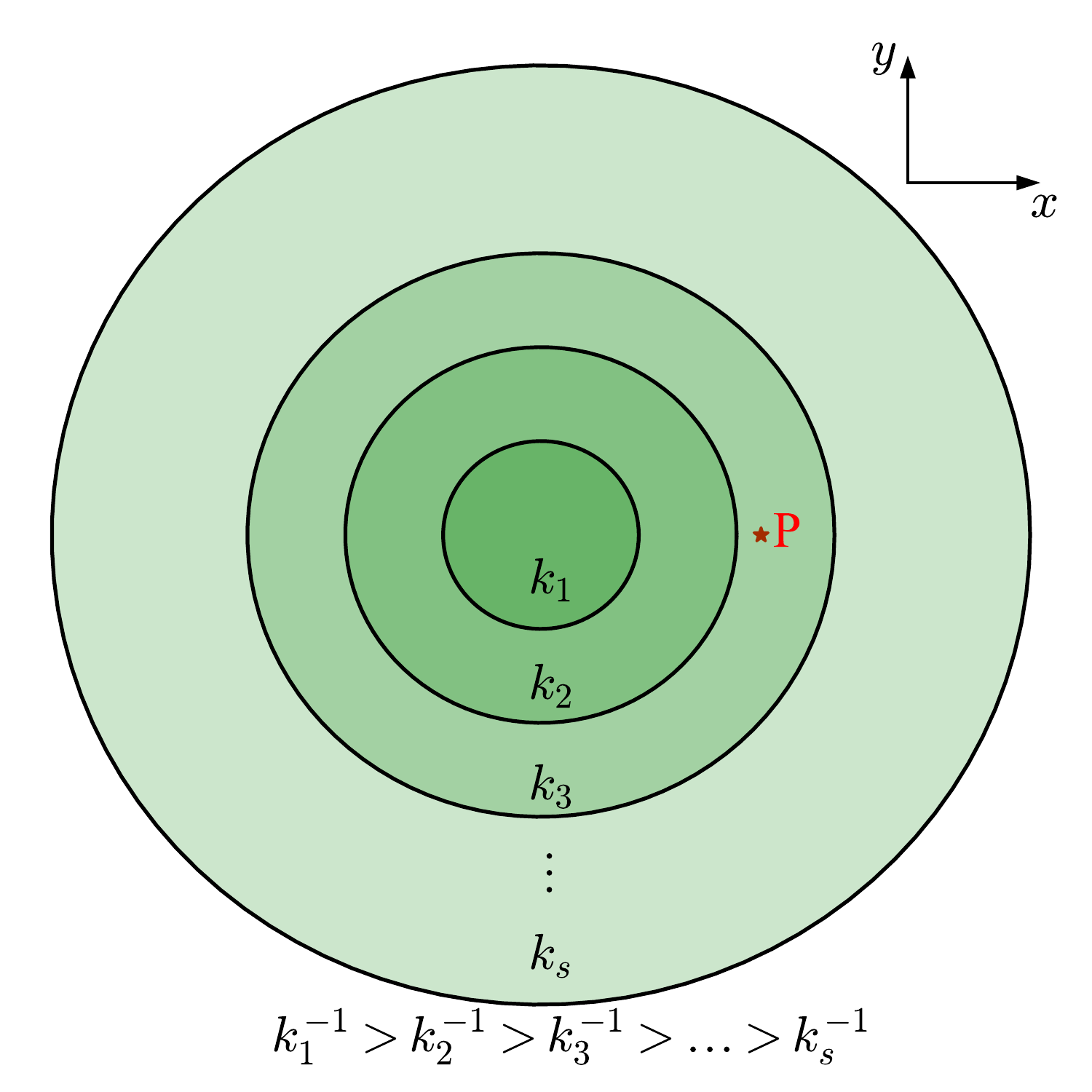}
	\caption{Concentric fuzzy spheres with the same $N$, but effective $k$ varies with the radius. Note that $k^{-1}$ behaves as the strength of NC in a squashed fuzzy sphere, which decreases from the center to the surface.}
	\label{Fig: Concentric spheres}
\end{figure}

Equation~\eqref{Eq: Fuzzy dispersion relation} provides the energy dispersion relation of one squashed fuzzy sphere corresponding to the wavefunctions of the degenerate elctrons, inside which the strength of NC ($k^{-1}$) is constant. Let us now consider a series of concentric fuzzy spheres (each of them are squashed in a common equatorial plane) with the same values of $N$. This means that all these fuzzy spheres can be explained by the same $\mathtt{SU}(2)$ algebra. Moreover, from Equation~\eqref{Eq: k_R relation}, for a fixed $N$, we have $k\propto r^2$, i.e., $k$ increases from the center to the surface; therefore, the strength of NC decreases from center to the surface. Hence the effective NC at a point of radius $r$, is due to the contributions from all the concentric spheres with radius more than $r$. As a result, effective $k$ no longer remains constant inside the sphere, and the variation of $k$ needs to be chosen so that $k$ is minimum at the center and maximum at the surface. Figure~\ref{Fig: Concentric spheres} shows an illustrative diagram of the variation of $k$ inside a fuzzy sphere. The effective NC at a point P is due to the contributions of NC from all the fuzzy spheres within which the point is enclosed. Since $k^{-1}$ behaves as the strength of NC in a squashed fuzzy sphere, effective NC is maximum at the center, which gradually decreases outwards and becomes minimum (mathematically zero) at the surface.

\section{Degenerate equation of state in squashed fuzzy sphere}\label{Degenerate equation of state in squashed fuzzy sphere}

\begin{figure}[htp]
	\centering
	\includegraphics[scale=0.5]{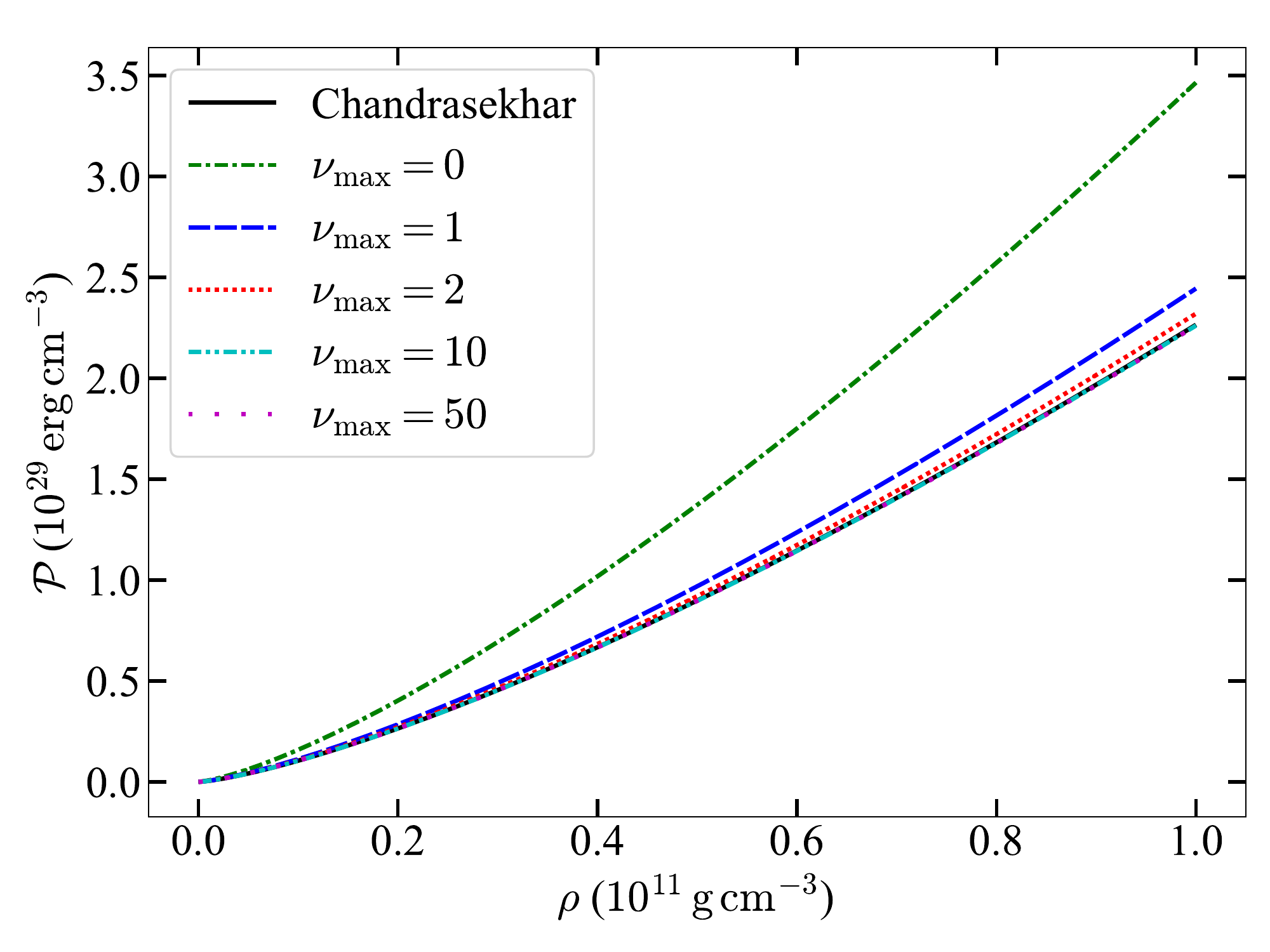}
	\caption{EoS for degenerate electrons in a squashed fuzzy sphere with various $\nu_\mathrm{max}$ along with the Chandrasekhar EoS. $\nu_\mathrm{max}=0$ means electrons occupy only the ground energy level, $\nu_\mathrm{max}=1$ means they occupy both ground and first energy levels, and so on. Chandrasekhar EoS corresponds to $\nu_\mathrm{max}\to\infty$.}
	\label{Fig: Fuzzy EoS}
\end{figure}

Since in this paper, our primary focus is to apply the effect of NC in the squashed fuzzy sphere for WDs, we now discuss the EoS of degenerate electrons for this case. The grand canonical potential for this system is given by \cite{1969stph.book.....L}
\begin{align}
\Omega &= -\frac{4V}{\beta k h^2} \int_{0}^{p_{r,\mathrm{max}}} \dd{p_r} \sum_{\nu=0}^{\nu_\mathrm{max}} g_\nu \ln\left(1+ e^{\beta(\mu-E)}\right) \\
&= -\frac{4V}{\beta k h^2} \int_{0}^{p_{r,\mathrm{max}}} \dd{p_r} \sum_{\nu=0}^{\nu_\mathrm{max}} g_\nu \ln\left(1+ ze^{-\beta E}\right), \label{Eq: Grand canonical potential}
\end{align}
where $z = e^{\beta \mu}$ is the fugacity, $\beta = 1/k_B T$, $\mu$ is the chemical potential, $V$ is the volume of the system, $k_B$ is the Boltzmann constant, $T$ is the temperature, and $g_\nu$ is the degeneracy factor with $g_\nu = 2 - \delta_{\nu 0}$ where $\delta_{\mu\nu}$ being the Kronecker delta function. Note that $V$ is the noncommutative volume, which is different from the classical volume. 
This is because in the commutative limit, the classical volume element is obtained as $\dd V = r^2 \sin\theta \dd r \dd\theta \dd\phi$, which may not hold good in the case of NC as $\theta$- and $\phi$-coordinates do not commute with each other. The relation between $\Omega$ and the number density of electrons ($n_\mathrm{e}$) is given by
\begin{align}\label{Eq: number density}
n_\mathrm{e} = -\frac{\beta z}{V} \pdv{\Omega}{z}.
\end{align}
Hence, the mass density ($\rho$) is given by
\begin{align}\label{Eq: mass density}
\rho = m_\mathrm{p} \mu_\mathrm{e} n_\mathrm{e},
\end{align}
where $m_\mathrm{p}$ is the mass of a proton and $\mu_\mathrm{e}$ is the mean molecular weight per electron. Since in WDs, $T \ll T_\mathrm{F}$ or $E\ll E_\mathrm{F}$ with $T_\mathrm{F}$ and $E_\mathrm{F}$ respectively being the Fermi temperature and Fermi energy, they are usually assumed to be cold, and the electrons become degenerate. For cold WDs with $T \ll T_\mathrm{F}$, we have $\mu \approx E_\mathrm{F}$. With this approximation in the large thermodynamic limit, using equations~\eqref{Eq: Grand canonical potential}, \eqref{Eq: number density} and \eqref{Eq: mass density}, we obtain
\begin{align} \label{Eq: Fuzzy density}
\rho &\approx \frac{4m_\mathrm{p} \mu_\mathrm{e} p_\mathrm{F}}{k h^2} \sum_{\nu=0}^{\nu_\mathrm{max}} g_\nu = \frac{4m_\mathrm{p} \mu_\mathrm{e} p_\mathrm{F}}{k h^2} \left(2\nu_\mathrm{max}+1\right),
\end{align}
where $p_\mathrm{F}$ is the Fermi momentum of the degenerate electrons in the squashed fuzzy sphere. Here $\nu_\mathrm{max}$ is the quantity that carries the information of the number of occupied energy levels. Now, defining $\theta_\mathrm{D} = 2 \hbar/m_\mathrm{e}^2 c^2 k$ and using Equation~\eqref{Eq: Fuzzy dispersion relation}, the relation between $E_\mathrm{F}$ and $p_\mathrm{F}$ is given by
\begin{align}
E_\mathrm{F}^2(\rho,\nu) = p_\mathrm{F}^2(\rho)c^2 + m_\mathrm{e}^2c^4\left\{1 + 2\nu \theta_\mathrm{D}(\rho)\right\}.
\end{align}
Since $p_\mathrm{F}^2\geq0$, using this equation, we obtain $E_\mathrm{F}^2 \geq m_\mathrm{e}^2c^4\left(1 + 2\nu \theta_\mathrm{D}\right) \implies \nu \leq \left(E_\mathrm{F}^2 - m_\mathrm{e}^2c^4\right)/2m_\mathrm{e}^2c^4\theta_\mathrm{D}$. Hence, $\nu_\mathrm{max}$ is defined as the greatest integer less than or equal to $\left(E_\mathrm{F,max}^2 - m_\mathrm{e}^2c^4\right)/2m_\mathrm{e}^2c^4\theta_\mathrm{D}$, i.e.
\begin{align}
\nu_\mathrm{max} &= \left[\frac{E_{\mathrm{F,max}}^2 - m_\mathrm{e}^2c^4}{2m_\mathrm{e}^2c^4\theta_\mathrm{D}} \right] \\
\text{or, } E_\mathrm{F,max}^2 &= m_\mathrm{e}^2 c^4 \left(1+2\nu_\mathrm{max}\theta_\mathrm{D} \right), 
\end{align}
where $[\vdot]$ is the floor function. Let us now assume that $\theta_\mathrm{D} \propto \rho^{2/3}$, such that the variation of $k$ with respect to $\rho$ is proposed to be
\begin{align}\label{Eq: k_rho relation}
k = \xi\frac{\mu_\mathrm{e}^{2/3} m_\mathrm{p}^{2/3}}{h \rho^{2/3}},
\end{align}
where $\xi$ is a dimensionless proportionality constant. The power of $\rho$ (which we assume to be $2/3$) is chosen in such a way that at the low-density limit, the pressure-density relation follows the Chandrasekhar EoS for degenerate electrons. This equation provides the variation of $k$ with $\rho$, as shown in Figure~\ref{Fig: Concentric spheres}. Since $k^{-1}\propto \rho^{2/3}$, which means $k$ is minimum at the center of the star where the density is maximum; it implies that the NC is maximum at the center. Towards the surface, density decreases gradually and, hence, $k$ increases accordingly, producing a minimal NC towards the surface. Substituting this expression for $k$ in Equation~\eqref{Eq: Fuzzy density}, we obtain
\begin{align}\label{Eq: p_F rho relation}
p_\mathrm{F} = \frac{\xi h}{4 \mu_\mathrm{e}^{1/3} m_\mathrm{p}^{1/3} \left(2\nu_\mathrm{max}+1\right)} \rho^{1/3}.
\end{align}
From this equation, we obtain $p_\mathrm{F} \propto \rho^{1/3}$, which Chandrasekhar also obtained in his theory \cite{1935MNRAS..95..207C} except that the proportionality constant is now different. Moreover, in the large thermodynamic limit (i.e., $E\ll E_\mathrm{F}$), the pressure ($\mathcal{P}$) is given by \cite{1991ApJ...383..745L}
\begin{align}\label{Eq: Fuzzy pressure}
\mathcal{P} &= \frac{2}{k h^2} \sum_{\nu=0}^{\nu_\mathrm{max}} g_\nu \left\{ p_\mathrm{F} E_\mathrm{F} - \left(m_\mathrm{e}^2 c^3 +2\nu \frac{2\hbar c}{k}\right) \ln(\frac{E_\mathrm{F} + p_\mathrm{F}c}{\sqrt{m_\mathrm{e}^2c^4 + 2\nu \frac{h c^2}{\pi k}}}) \right\} \nonumber \\
&= \frac{2 \rho^{2/3}}{\xi h \mu_\mathrm{e}^{2/3} m_\mathrm{p}^{2/3}} \sum_{\nu=0}^{\nu_\mathrm{max}} g_\nu \left\{ p_\mathrm{F} E_\mathrm{F} - \left(m_\mathrm{e}^2 c^3 +2\nu \frac{2\hbar c}{k}\right) \ln(\frac{E_\mathrm{F} + p_\mathrm{F}c}{\sqrt{m_\mathrm{e}^2c^4 + 2\nu \frac{h c^2}{\pi k}}}) \right\}.
\end{align}
This equation for $\mathcal{P}$ combining with Equation~\eqref{Eq: p_F rho relation}, gives the complete EoS. Figure~\ref{Fig: Fuzzy EoS} shows the EoS for degenerate electrons in a squashed fuzzy sphere with various $\nu_\mathrm{max}$. The quantity $\xi$ for each $\nu_\mathrm{max}$ is chosen so that at low density, the EoS merge to the Chandrasekhar EoS. It is evident that as $\nu_\mathrm{max}$ increases, noncommutative EoS merges with the Chandrasekhar EoS.


\subsection{Scales of NC in squashed fuzzy sphere}\label{Scales of NC in squashed fuzzy sphere}
The structure of NC in a squashed fuzzy sphere is essential to study the physics of a stellar object as NC becomes zero at its surface. In other words, since the density is minimal (mathematically zero) at the surface, NC effect should vanish, and the usual physics given by the QM must be restored there. Moreover, as mentioned above, $k^{-1}$ is the strength of NC in a squashed fuzzy sphere. From Equation~\eqref{Eq: k_R relation}, it is evident that $r \propto \sqrt{k}$, which means that fuzziness is less at a bigger radius, and the usual QM dominates the underlying statistical mechanics. As a result, WDs with less central densities (which usually have bigger radii) are expected to possess a larger value of $k$ at the center than those with higher densities (or smaller radii).

We know that in the presence of a magnetic field, the effect of Landau quantization is prominent if $B\gtrsim B_\mathrm{c}$, which is evident from Equation~\eqref{Eq: Magnetic EoS}. Let us now examine the equivalent condition for the squashed fuzzy sphere algebra. Equation~\eqref{Eq: Fuzzy dispersion relation} can be recast to
\begin{align}
E^2 = p_r^2 c^2 + m_\mathrm{e}^2 c^4 \left(1+ 2\nu \frac{\lambda_\mathrm{e}^2}{2 \pi^2 \hbar k}\right),\qquad \nu \in \mathbb{Z}^{0+}
\end{align}
where $\lambda_\mathrm{e}$ is the Compton wavelength of an electron with $\lambda_\mathrm{e} = h/m_\mathrm{e}c \approx 2.426 \times 10^{-10}\rm\, cm$. Hence, the effect of NC is noticeable if $\lambda_\mathrm{e}^2\gtrsim2 \pi^2 \hbar k$. Moreover, we know that the characteristic length-scale of a system is the inter-particle separation, given by $\mathfrak{L} = (\rho/\mu_\mathrm{e}m_\mathrm{p})^{-1/3}$. Substituting $k$ as a function of $\rho$ from Equation~\eqref{Eq: k_rho relation}, we obtain the following relation
\begin{align}\label{Eq: Length scale}
\mathfrak{L} \lesssim \frac{\lambda_\mathrm{e}}{\sqrt{\pi \xi}} = \mathfrak{L}_\mathrm{eff}.
\end{align}
This is the condition for which the effect of NC to be significant in the case of degenerate electrons. It is evident from this equation that the effective length-scale for the prominence of NC is not only dependent on the Planck length, but it depends upon the system's scale also. This idea of length-scale can also be followed from the idea of Salecker and Wigner \cite{1958PhRv..109..571S}, which states that the new uncertainty in length-scale for a system has to be $\delta \sim \left(\mathfrak{L} \mathfrak{L}_\mathrm{P}^2\right)^{1/3}$, where $\mathfrak{L}_\mathrm{P}$ is the Planck length, and one needs to consider $\delta$ as the quantum measurement of length. Hence, for a WD, where $\mathfrak{L}\gg\mathfrak{L}_\mathrm{P}$, the effective scale of NC turns out to be $\delta\gg\mathfrak{L}_\mathrm{P}$. As a result, one can expect to observe a significant NC effect in a WD, even though the system's scale is far from the Planck scale. In the next section, we show that the effect of NC is prominent only in the highly-dense regime, whereas its effect is insignificant at the low-density and, hence, this formalism does not violate any of the observables in the low-density universe, such as the solar system.

This idea of NC's prominence depends on the system's length-scale reasonably overcomes one of the puzzles lying with the magnetic field. In the case of a magnetic field, as mentioned above, its quantum effect is prominent only if $B\gtrsim B_\mathrm{c} = 4.414 \times 10^{13}\rm\, G$. However, we know that the magnetic field has both classical and quantum effects. Due to its classical effect, the magnetic field can change the WD's shape and size, thereby increasing its mass too. If a WD possesses a field significantly stronger than $B_\mathrm{c}$, such a high field may also destabilize the WD, depending on the geometry, because of its high magnetic to gravitational energy ratio. Hence, the existence of a high field inside a WD is debatable. This problem can consistently be sorted out in the case of a noncommutative squashed fuzzy sphere. In our formalism of a squashed fuzzy sphere, since the NC is between the azimuthal and polar directions and it does not have any classical effect, the system's sphericity is not destroyed. Moreover, the length-scale for which NC has a noticeable effect is set-in by the theory itself and we do not need to assume an ad-hoc length-scale beforehand. Hence any further physical effect of NC will turn out to be an emergent phenomenon.

\section{Limiting mass of the WD in squashed fuzzy sphere}\label{Limiting mass of the WD in squashed fuzzy sphere}

\begin{figure}[htp]
	\centering
	\includegraphics[scale=0.5]{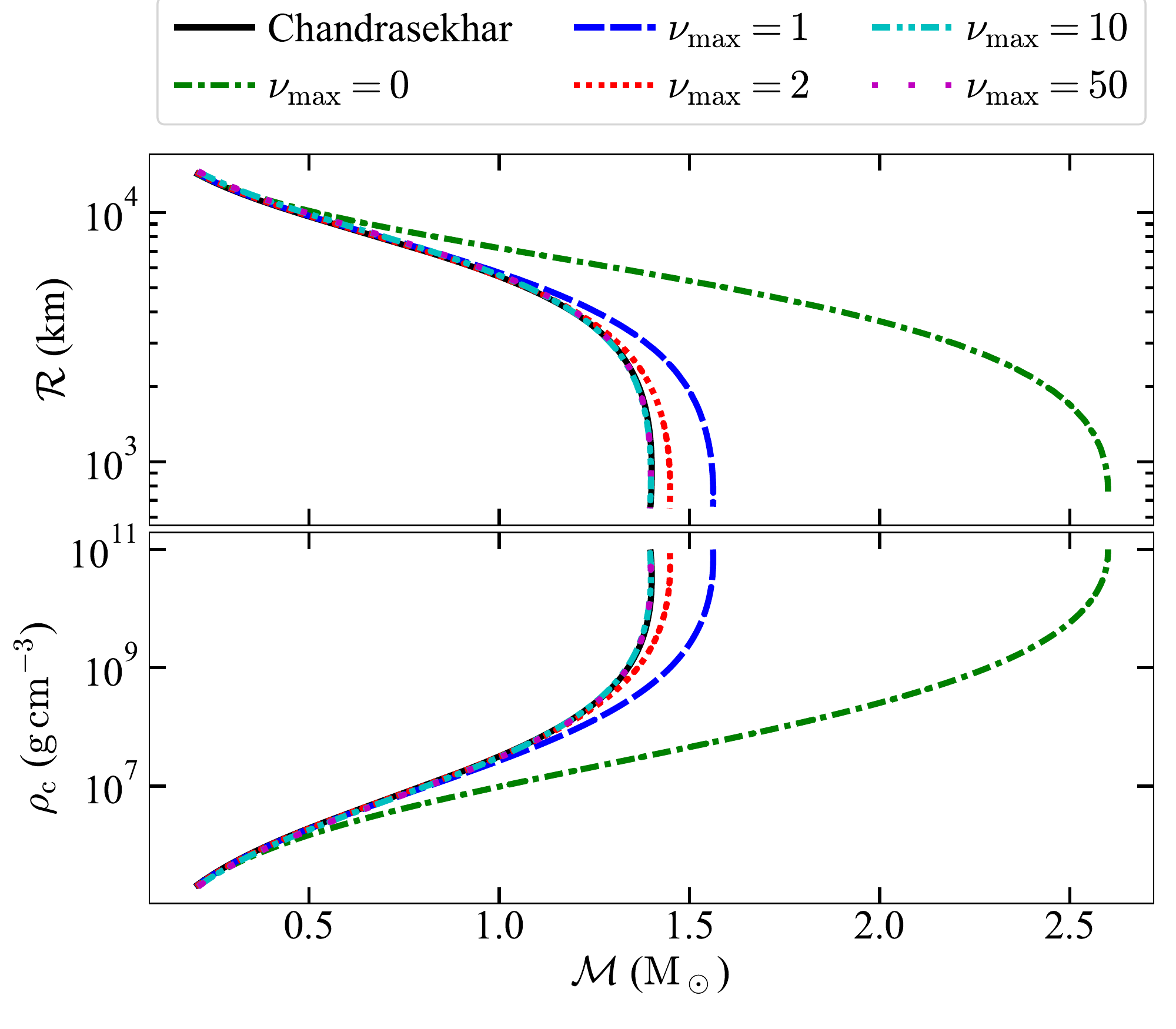}
	\caption{Upper panel: The mass--radius relation, Lower panel: The variation of central density with the mass of the WD.}
	\label{Fig: Fuzzy M_R}
\end{figure}

The hydrostatic balance of any stellar object is obtained by simultaneously solving the pressure balance and mass balance equations (together known as the Tolman-Oppenheimer-Volkoff or TOV equations) along-with the EoS of the constituent particles. The TOV equations are given by \cite{2009igr..book.....R}
\begin{equation}\label{TOV}
\begin{aligned}
\dv{\mathcal{M}}{r} &= 4\pi r^2\rho,\\
\dv{\mathcal{P}}{r} &= -\frac{G}{r^2}\left(\rho+\frac{\mathcal{P}}{c^2} \right)\left(\mathcal{M}+\frac{4\pi r^3 \mathcal{P}}{c^2}\right) \left(1-\frac{2G\mathcal{M}}{c^2r}\right)^{-1},
\end{aligned}
\end{equation}
where $\mathcal{M}$ is the mass of the star inside a volume of radius $r$ and $G$ is Newton's gravitational constant. In our simplictic assumption, NC affects the microscopic physics, while TOV equations describe the macroscopic physics. Hence, the treatment is semiclassical. Therefore, the TOV equations remain the same as that for classical general relativistic formalism, where electrons play the role of giving pressure to compensate the gravitational pull, whereas mass mostly is determined by the other baryons. Figure~\ref{Fig: Fuzzy M_R} shows the mass--radius relation as well as the variation of the mass with respect to the central density $\rho_\mathrm{c}$ for the WDs which is obtained by solving the set of TOV equations~\eqref{TOV} along with the EoS obtained in the previous section. For comparison, we also present Chandrasekhar's results therein. The new mass-limit is the consequence of the prominence of NC when the inter-electron distance $\mathfrak{L}\lesssim\mathfrak{L}_\mathrm{eff}$. As mentioned in the \S \ref{Degenerate equation of state in squashed fuzzy sphere}, it is important that the effect of NC should be insignificant at lower densities and the usual statistical mechanics, governed by the usual QM, should prevail there.

From Figure~\ref{Fig: Fuzzy M_R}, it is evident that as the occupied energy levels (i.e., $\nu_\mathrm{max}$) increases, the effect of NC decreases, and eventually all the curves merge with the Chandrasekhar mass--radius curve. It can easily be understood from the simple length-scale idea mentioned in \S \ref{Scales of NC in squashed fuzzy sphere}. To match our EoS at low-density with Chandrasekhar's, we find $\xi=1.51$ for $\nu_\mathrm{max}=0$ while $\xi=156.56$ for $\nu_\mathrm{max}=50$. Substituting these values in the relation \eqref{Eq: Length scale}, we find that the NC is prominent if $\mathfrak{L}\lesssim 1.11\times10^{-10}\rm\, cm$ for $\nu_\mathrm{max}=0$ whereas $\mathfrak{L}\lesssim 1.09\times10^{-11}\rm\, cm$ for $\nu_\mathrm{max}=50$. These length-scales correspond to $\rho \sim 2.4 \times 10^6\rm\, g\, cm^{-3}$ and $\rho \sim 2.6\times 10^9\rm\, g\, cm^{-3}$. Hence, for $\nu_\mathrm{max}=0$, the effect of NC is significant in the regime of the WD with $\rho \gtrsim 2.4\times 10^6\rm\, g\, cm^{-3}$, while the corresponding $\rho \gtrsim 2.6\times 10^9\rm\, g\, cm^{-3}$ for $\nu_\mathrm{max}=50$. As a result, we do not observe any significant effect of NC in the WDs in the allowed density regime for $\nu_\mathrm{max}=50$, and it continues to follow the Chandrasekhar's original mass--radius relation. Moreover, from equations~\eqref{Eq: k_R relation} and \eqref{Eq: k_rho relation}, we obtain $N\sim 10^{37}-10^{40}$ for different $\nu_\mathrm{max}$, which verifies the assumption of large $N$ in Equation~\eqref{Eq: Fuzzy dispersion relation} does hold good.

\section{Neutron drip in the presence of noncommutativity}\label{Neutron drip in presence of noncommutativity}

We have obtained the modified dispersion relation in Equation~\eqref{Eq: Fuzzy dispersion relation} for a squashed fuzzy sphere. Since this is a different dispersion relation compared to that of the relativistic case, one can expect that the property of neutron drip alters in the presence of NC. A detailed discussion of neutron drip in the presence of a magnetic field is given by Vishal and Mukhopadhyay \cite{2014PhRvC..89f5804V}. At the neutron drip density, protons and electrons combine to form neutrons, and they come out of the nucleus to form a neutron lattice. As a result, above this density, the electron degenerate pressure is no longer available, and the EoS obtained in \S \ref{Degenerate equation of state in squashed fuzzy sphere} is no longer valid. The neutron drip density $\rho_\text{drip}$ is given by
\begin{equation}
\rho_\text{drip} = \frac{n_\mathrm{e} M(A,Z)/Z + \epsilon_\mathrm{e}- n_\mathrm{e} m_\mathrm{e} c^2}{c^2},
\end{equation}
where $M(A,Z)$ is the energy of a single ion with atomic number $Z$ and mass number $A$. $\epsilon_\mathrm{e}$ is the electron energy density at zero temperature, which is given by \cite{1991ApJ...383..745L}
\begin{align}
\epsilon_\mathrm{e} = m_\mathrm{e} c^2 \frac{4\pi\theta_\mathrm{D}}{\lambda_\mathrm{e}^3} \sum_{\nu=0}^{\nu_\mathrm{max}}g_{\nu}\left(1+\nu \theta_\mathrm{D}\right) \Psi\left(\frac{x_\mathrm{F}(\nu)}{\sqrt{1+2\nu \theta_\mathrm{D}}}\right),
\end{align}
where
\begin{align*}
x_\mathrm{F} &= \left(2230.31-2\nu \theta_\mathrm{D}\right)^{1/2}, \quad
\Psi(z) = \frac{1}{2}z\sqrt{1+z^2}+\frac{1}{2}\ln \left(z+\sqrt{1+z^2}\right).
\end{align*}
\begin{figure}[htpb]
	\centering
	\includegraphics[scale=0.6]{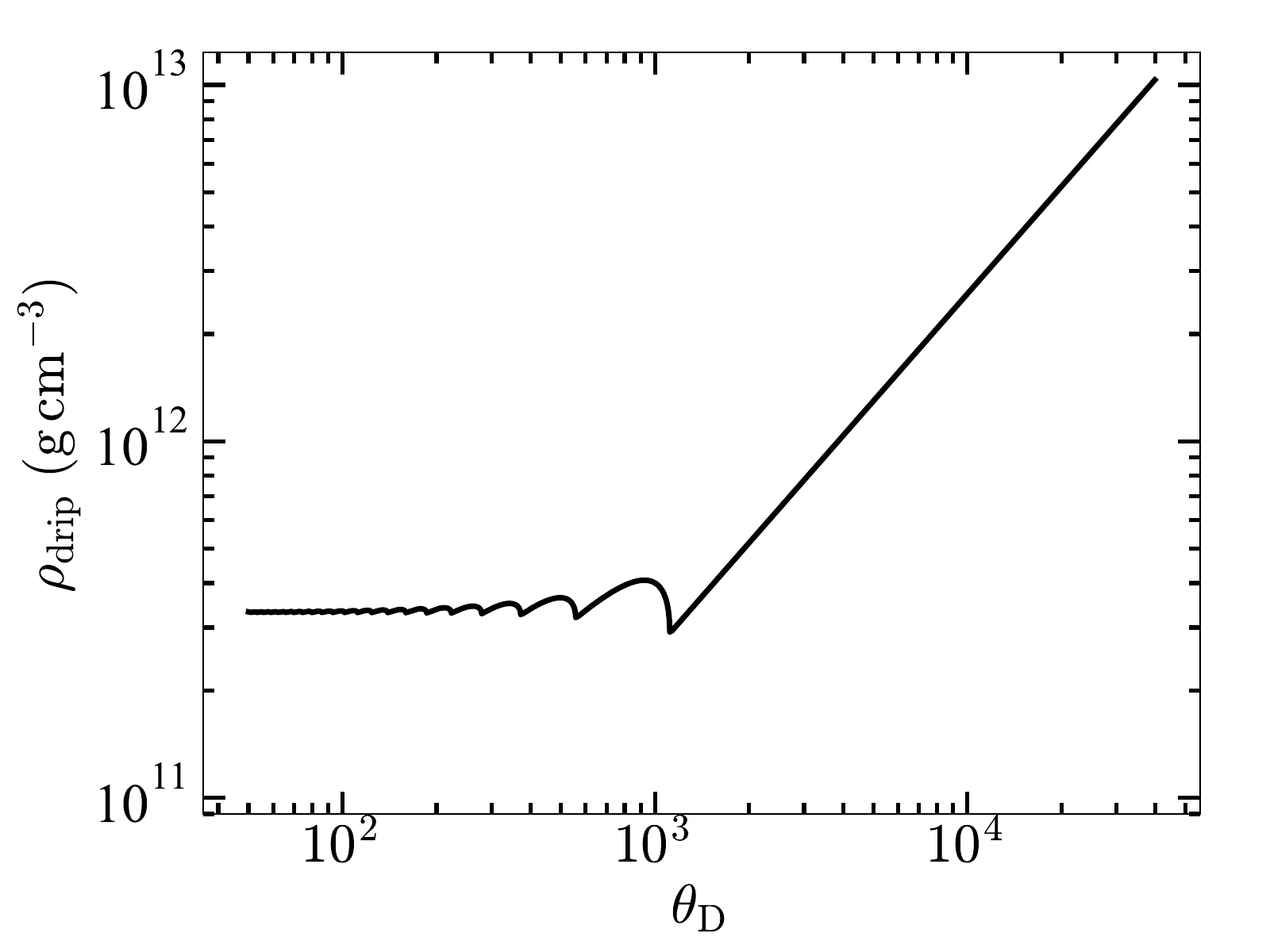}
	\caption{Variation of neutron drip density as a function of $\theta_\mathrm{D}$.}
	\label{Fig: density_drip}
\end{figure}

Figure~\ref{Fig: density_drip} shows the variation of $\rho_\text{drip}$ with respect to $\theta_\mathrm{D}$. We notice that for $\theta_\mathrm{D}$ below the linear regime, the drip density oscillates about the density obtained in the absence of NC which is $\sim 3 \times 10^{11}\rm\, g\, cm^{-3}$. It is also evident from the figure that the onset drip density increases linearly above a certain $\theta_\mathrm{D}$. This is because, beyond certain $\theta_\mathrm{D}$, the quantized electrons reside only in the ground state. To summarize, $\rho_\text{drip}$ changes with the increase in NC. Above this density, various nuclear reactions occur, and we can no longer use the EoS of non-interacting degenerate electrons obtained in \S \ref{Degenerate equation of state in squashed fuzzy sphere}. Such nuclear reactions are usually prominent at the core of neutron stars. If NC also prevails there, provided the inter-particle separation is less than the constituent particles' Compton wavelengths, it can also increase the mass of a neutron star. Such a massive neutron star can possibly explain the existence of $2.6\rm\,M_\odot$ compact object in the GW190814 merger event \cite{2020ApJ...896L..44A}. However, applying NC self-consistently in neutron stars is beyond the scope of this work.

\section{Conclusions}\label{Conclusions}

NC is a fundamental property of matter and spacetime geometry. In the presence of NC, the statistical behaviour of the constituent particles changes. Because of this, in NC, the fermions behave less of conventional fermions. It is evident from Figure~\ref{Fig: Total energy levels} that apart from the ground level, each energy levels comprise of two up-spin and two down-spin electrons. In the conventional picture of QM, we know that two electrons with the same spin cannot occupy a common energy state due to the Pauli's exclusion principle. However, since more than one electrons with the same spin occupy the common energy state in squashed fuzzy sphere NC, the effective repulsive pressure between the fermions decreases. Moreover, the Chandrasekhar mass-limit is the consequence of the competition between effective fermion pressure and gravitational attraction. If a fermion behaves less like a fermion (more like a boson), the inter-particle separation decreases, which generates a higher matter density tempting to accumulate more mass. Moreover, when the central density becomes $\sim 10^{8-10}\rm\, g\, cm^{-3}$, the inter-electron separation is less than the Compton wavelength of electrons, which triggers NC, and the core pressure increases to support more mass. However, fermion, losing its identity, leads to less repulsion among themselves, which further leads to less outward pressure to hold more mass. Combining all three effects, nevertheless, we obtain that the hydrostatic equilibrium is obtained at a higher mass for the highly dense WDs. Hence, they can collapse to a smaller size and extra mass can be accumulated to form super-Chandrasekhar WDs. In other words, in the presence of NC, for a given density, the degeneracy pressure is higher as shown in Figure~\ref{Fig: Fuzzy EoS}. Hence, hydrostatic equilibrium can be maintained to higher masses, preventing collapse.

We previously investigated the effect of NC in a WD with the ad-hoc consideration of $\comm{\hat{x}_i}{\hat{x}_j} = i\eta_{ij}$ and $\comm{\hat{p}_i}{\hat{p}_j} = i\theta_{ij}$, and showed that it could lead to a super-Chandrasekhar WD if we assume that the NC is effective if the length-scale of the system is less than the Compton wavelength of electrons \cite{2021IJMPD..3050034K}. However, in the present exploration of NC through the squashed fuzzy sphere, the effect of NC's prominence is an emergent quantity through Equation~\eqref{Eq: Length scale}, and there is no need to pre-assume any ad-hoc length-scale. It is essential to mention that our work is primarily based on semi-classical gravity, not quantum gravity and, hence, the scale of NC is expected to be different from that of the Planck scale. In the context of specific dynamics of quantum mechanics, gravity, and statistical mechanics, the NC scale gets complicated. From Equation~\eqref{Eq: Length scale}, it is evident that if $\mathfrak{L}<\mathfrak{L}_\mathrm{eff}$, the NC is efficient. Indeed, Salecker and Wigner already pointed it out as a quantum measurement of lengths \cite{1958PhRv..109..571S}. Hence, the new uncertainty in length-scale appears to be $\delta \sim \left(\mathfrak{L} \mathfrak{L}_\mathrm{P}^2\right)^{1/3}$. To summarize, the length-scale of uncertainty in a WD, which is a low energy system, depends both on the Planck length and length-scale of the system. 
In this article, we have modeled WDs with noncommutating degenerate electrons in 3-dimensions, in such a way that NC appears self-consistently. The idea of the collection of fuzzy spheres, like an onion, is a suitable model, which was earlier used for modeling black holes. However, unfortunately, it does not mimic the reality as NC is the same throughout, whereas we expect it to vanish when the density becomes zero. Hence, the squashed fuzzy sphere is a good alternative that produces a reasonable noncommutative geometry. We have modeled the WDs in such a way that NC decreases from the center to the surface and becomes zero at the boundary of the WDs. NC on each shell is between the polar and azimuthal coordinates, which need to be obtained from the projection. These quantum states are discrete, but we cannot think of them in terms of electrons in a lattice. This is similar to the Moyal plane, where positions $x$ and $y$ alone do not make sense; instead, their expectation values are helpful. So we should view the configuration as a quantum spherical ball. Note that NC, being a quantum effect, does not affect the WD's classical spherical geometry.

Earlier explorations for the formation of super-Chandrasekhar WDs in high magnetic fields and thereby the Landau levels was an interesting idea. However, we know that the magnetic field has not only microscopic effects but also macroscopic effects. As a result, in the presence of high varying magnetic fields, a WD would be deformed due to the Lorentz force of the magnetic field. Various explorations set limits on the maximum magnetic to gravitational energy ratio in a compact object \cite{2009MNRAS.397..763B}. Hence, the presence of a very high magnetic field inside a compact object is debatable. The idea of NC through the squashed fuzzy sphere fairly overcomes this problem as the fuzzy algebra's formalism is such that it does not affect the sphericity of the system. Note that our model Hamiltonian does not have the rotational invariance. However, in principle, one can restore the rotational symmetry because the projection can be along any of the equatorial planes. The additional effect of NC appears only through the change in pressure of degenerate electron gas at high densities. Moreover, from Equation~\eqref{Eq: Squashed fuzzy sphere}, the structure of NC is such that at the surface, the NC is zero. Hence, the effective NC varies with the density (or, equivalently with distance from the center), and its effect is prominent if the inter-particle separation $\mathfrak{L}<\mathfrak{L}_\mathrm{eff}$. Therefore, NC seems to be a better bet to explain the super-Chandrasekhar WDs as it does not possess any classical effect.

All the inferences of the super-Chandrasekhar WDs so far have been made indirectly, as none of them has been detected from direct observations. Similarly, there is no direct observational evidence for NC so far. Gravitational wave can be one of the prominent tools to detect such massive WDs directly. We earlier showed that WDs governed by noncommutative geometry can emit gravitational radiation for a long duration if they possess a minimal surface magnetic field \cite{2020ApJ...896...69K}. In this way, one can estimate both the masses and radii of the WDs; thereby comparing with the theoretical mass--radius curves, we might have indirect observational evidence for NC.

\section*{Acknowledgements}
SK would like to thank F. G. Scholtz of University of Stellenbosch and H. C. Steinacker of University of Vienna for their useful discussions on fuzzy sphere and underlying statistical mechanics during compilation of the work. The authors would also like to thank V. P. Nair of The City College of New York for his useful comments on noncommutative algebra.

\bibliographystyle{ws-ijmpd}
\bibliography{bibliography}

\end{document}